\documentclass[11pt]{article}
\usepackage{moriond,epsfig}
\usepackage{amsmath}

\def\Journal#1#2#3#4{{#1} {\bf #2}, #3 (#4)}

\def\PRD{{\em Phys. Rev.} D}

\def\be{\begin{equation}}
\def\ee{\end{equation}}
\def\bea{\begin{eqnarray}}
\def\eea{\end{eqnarray}}

\def\fig{\frenchspacing Fig.~}

\begin{document}
\vspace*{3cm}
\title{CONSTRAINTS ON THE TOPOLOGY OF THE UNIVERSE DERIVED FROM THE 7-YEAR WMAP CMB DATA AND PROSPECTS OF CONSTRAINING THE TOPOLOGY USING CMB POLARISATION MAPS}

\author{ P.~BIELEWICZ$^{1,2}$, A.J.~BANDAY$^{1,2}$, K.M.~G\'ORSKI$^{3,4}$ }

\address{$^1$ Universit\'e de Toulouse, UPS-OMP, Institut de Recherche en
  Astrophysique et Plan\'etologie, Toulouse, France \\ $^2$ CNRS, IRAP, 9 Av.~colonel
  Roche, BP 44346, F 31028 Toulouse cedex 4, France \\  $^3$ Jet
  Propulsion Laboratory, M/S 169/327, 4800 Oak Grove Drive, Pasadena
  CA 91109 \\ $^4$ Warsaw University Observatory, Aleje Ujazdowskie 4,
  00-478 Warszawa, Poland}

\maketitle
\abstracts{
We impose constraints on the topology of the Universe determined from
a search for matched circles in the cosmic
 microwave background (CMB) temperature anisotropy patterns of
the 7-year \emph{WMAP} data. We pay special attention to 
the sensitivity of the method to residual foreground contamination of
the sky maps. A search for pairs of matched back-to-back
circles in the higher resolution \emph{WMAP} W-band map allows tighter
constraints to be imposed on topology. Our results rule out universes with
topologies that predict pairs of such circles with radii larger than
$\alpha_{\rm min} \approx 10^\circ$. This places a lower bound on the
size of the fundamental domain for a flat universe of about 27.9 Gpc.
 We study also the possibility for constraining the topology of the Universe by
 means of the matched circles statistic applied to polarised CMB
 anisotropy maps. The advantages of using the CMB polarisation maps in
 studies of the topology over simply analysing the temperature data as has been done to-date
 are clearly demonstrated. It is found that the noise levels of both Planck and next 
 generation CMB experiments data are no longer prohibitive
 and should be low enough to enable the use of the
 polarisation maps for such studies.
}

\section{Introduction}
According to General Relativity, the pseudo-Riemannian manifold with signature (3,1) is
a mathematical model of spacetime. The local properties of spacetime geometry are described by the 
Einstein gravitational field equations. However, they do not specify
the global spatial geometry of the universe, i.e.\ its topology. This
can only be constrained by observations. The concordance cosmological model assumes that the universe 
possesses a simply-connected topology, yet various anomalies observed
on the largest angular scales in the \emph{WMAP} data in the last decade 
suggest that it may be multiply-connected. Evidence of such anomalies
comes from the suppression of the quadrupole moment and an alignment between the
preferred axes of the quadrupole and the octopole (Copi {\it et
  al}.~\cite{copi:2004}, de Oliveira-Costa {\it et al}.~\cite{de Oliveira-Costa:2004}).

We constrained the topology of the Universe (Bielewicz and
Banday~\cite{bielewicz:2011}) using the method of 
matched circles proposed by Cornish {\it et al}.~\cite{cornish:1998} and applied it to
the 7-year \emph{WMAP} data (Jarosik {\it et al}.~\cite{jarosik:2011}). In contrast to
the majority of previous studies, we paid special attention to the impact of
Galactic foreground residuals on the constraints. The 
method was applied to higher resolution maps than previously, which implies a
lower level of false detection and therefore tighter constraints on the size of the
Universe. As a result of computational limitations, we restricted the 
analysis to a search for back-to-back circle pairs\footnote{pairs of
  circles centred around antipodal points}. 

The method of matched circles is not inherently limited to temperature anisotropy
studies. It can also be applied to the CMB polarisation data
(Bielewicz {\it et al}.~\cite{bielewicz:2012}). We
investigate also such an application of the method. We 
test it on simulated CMB maps for a flat universe with the topology of  a
3-torus, and explicitly consider the possibility for the detection of matching circle pairs for data
with an angular resolution and noise level characteristic of the
Planck and COrE data. The latter is treated as a reference
mission for the next generation of CMB experiments.

\section{Statistic for the matched circles} \label{sec:statistic}
If light had sufficient time to cross the fundamental cell, an observer would see multiple copies of a single 
astronomical object. To have the best chance of seeing `around the universe' we should look for multiple 
images of distant objects. Searching for multiple images of the last scattering surface
is then a powerful way to constrain
topology. Because the surface of last scattering is a sphere centred on the observer, each copy of the observer
will come with a copy of the last scattering surface, and if the copies are separated by a distance less than the 
diameter of the last scattering surface, then they will intersect along 
circles. These are visible by both copies of the observer, but from opposite sides. The two copies are really 
one observer so if space is sufficiently small, the CMB radiation from the last scattering surface will contain
a pattern of hot and cold spots that match around the circles. 

The idea of using such circles to study topology
is due to Cornish {\it et al}.~\cite{cornish:1998}. Therein, a
statistical tool was developed to detect correlated circles in all sky maps of the CMB
anisotropy -- the circle comparison statistic
\begin{equation} \label{eqn:s_statistic}
S_{p,r}^{\pm} (\alpha, \phi_\ast)=\frac{\left<2\, X_p(\pm \phi) X_r(\phi+\phi_\ast)\right>}{\left<X_p(\phi)^2+X_r(\phi)^2\right>}\ ,
\end{equation}
where $\left< \right>=\int_0^{2\pi}d\phi$ and $X_p(\pm \phi)$,
$X_r(\phi+\phi_\ast)$ are temperature (or polarisation)
fluctuations around two circles of angular radius $\alpha$  centered
at different points, $p$ and $r$, on the sky with relative phase
$\phi_\ast$. The sign $\pm$ depends 
on whether the points along both circles are ordered in a clockwise direction
(phased, sign $+$) or alternately whether along one of the circles the
points are ordered in an anti-clockwise direction (anti-phased, sign
$-$). This allows the detection of both orientable and non-orientable topologies.
For orientable topologies the matched circles have anti-phased
correlations while for non-orientable
topologies they have a mixture of anti-phased and phased correlations.
To find correlated circles for each radius $\alpha$, the
maximum value $S_{\rm max}^{\pm}(\alpha) = \underset{p,r,\phi_\ast}{\rm max}
\, S_{p,r}^{\pm}(\alpha,\phi_\ast)$ is determined. In case of
anticorrelated circles the maximum value of
$-S_{p,r}^{\pm}(\alpha,\phi_\ast)$ is used. 
In the original paper by Cornish {\it et al}.~\cite{cornish:1998} the above statistic was
applied exclusively to temperature anisotropy maps. However, it can
also be applied to polarisation data and in this work, we focus on
its application to the E-mode map. In this case, the $X$ map is simply
the map of the E-mode.

To draw any conclusions from an analysis based on the statistic $S_{\rm
 max}^{\pm}(\alpha)$ it is important to correctly estimate the threshold
for a statistically significant match of the circle pairs. We
used simulations of the maps with the same noise properties and
smoothing scales as the data to establish the threshold
such that fewer than 1 in 100 simulations would yield a false event. 

\section{Constraints on the topology of the Universe derived from
  7-year WMAP data}\label{sec:wmap_results}
\begin{figure}
\centerline{
\epsfig{file=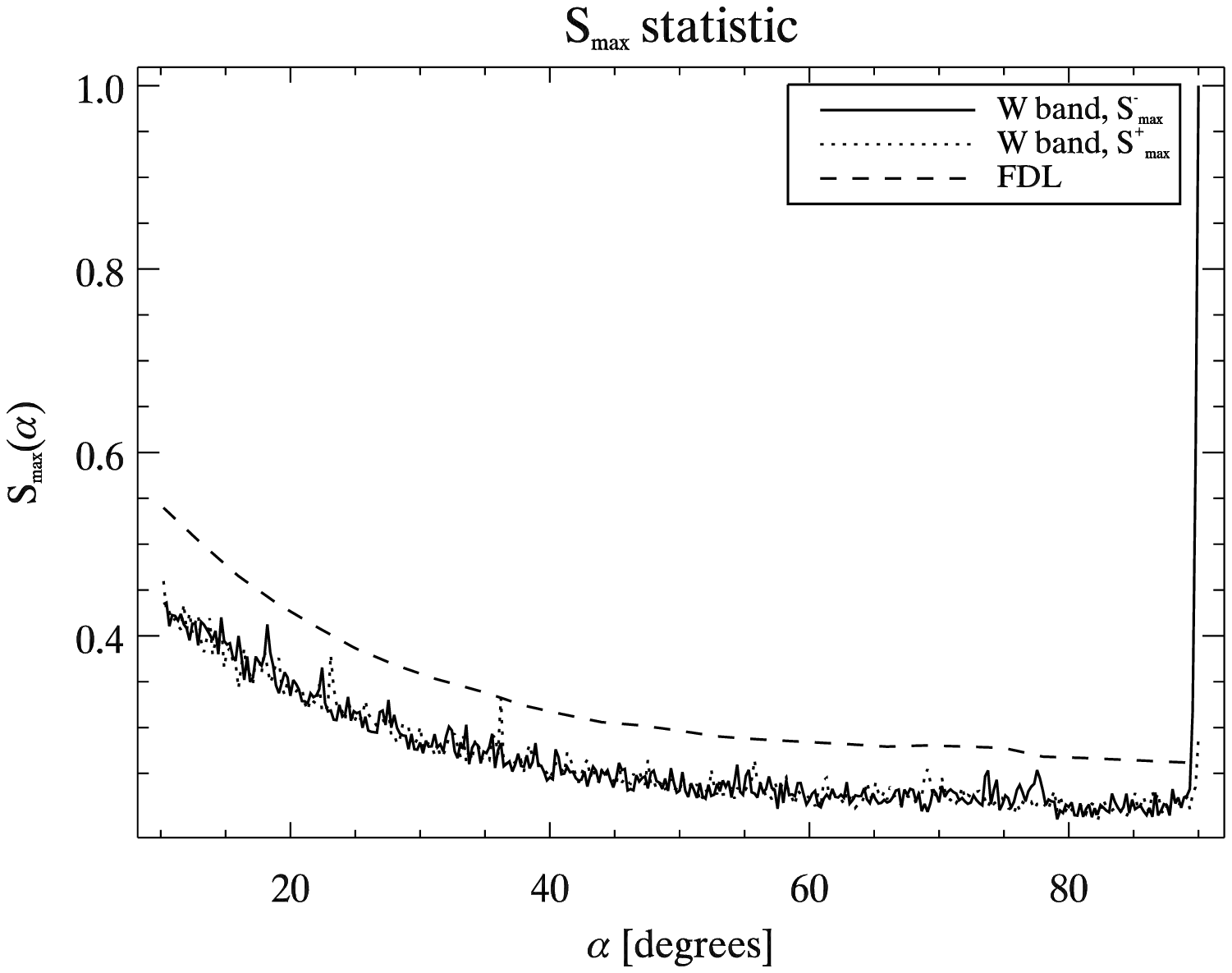,scale=0.35}
\epsfig{file=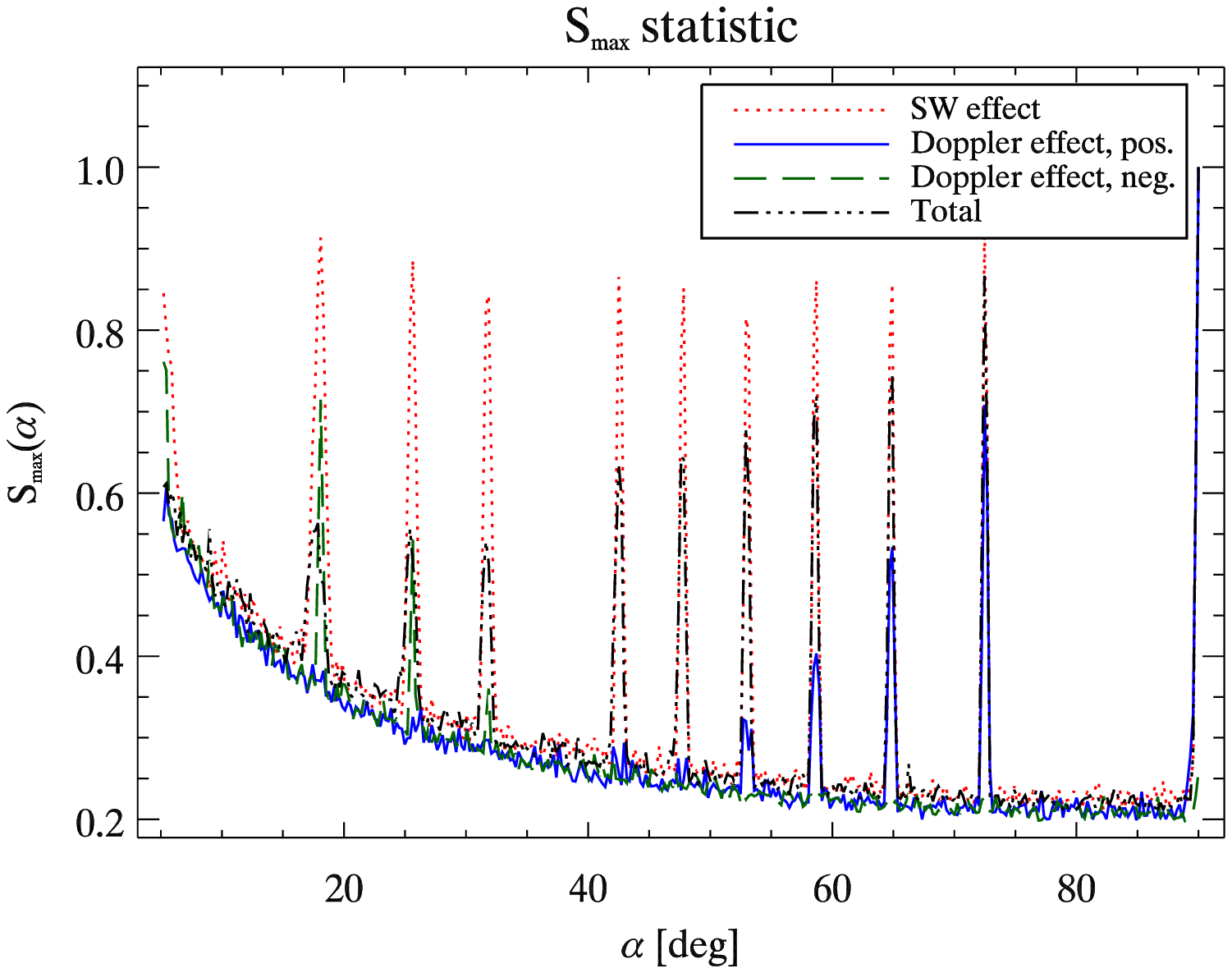,scale=0.35}
\epsfig{file=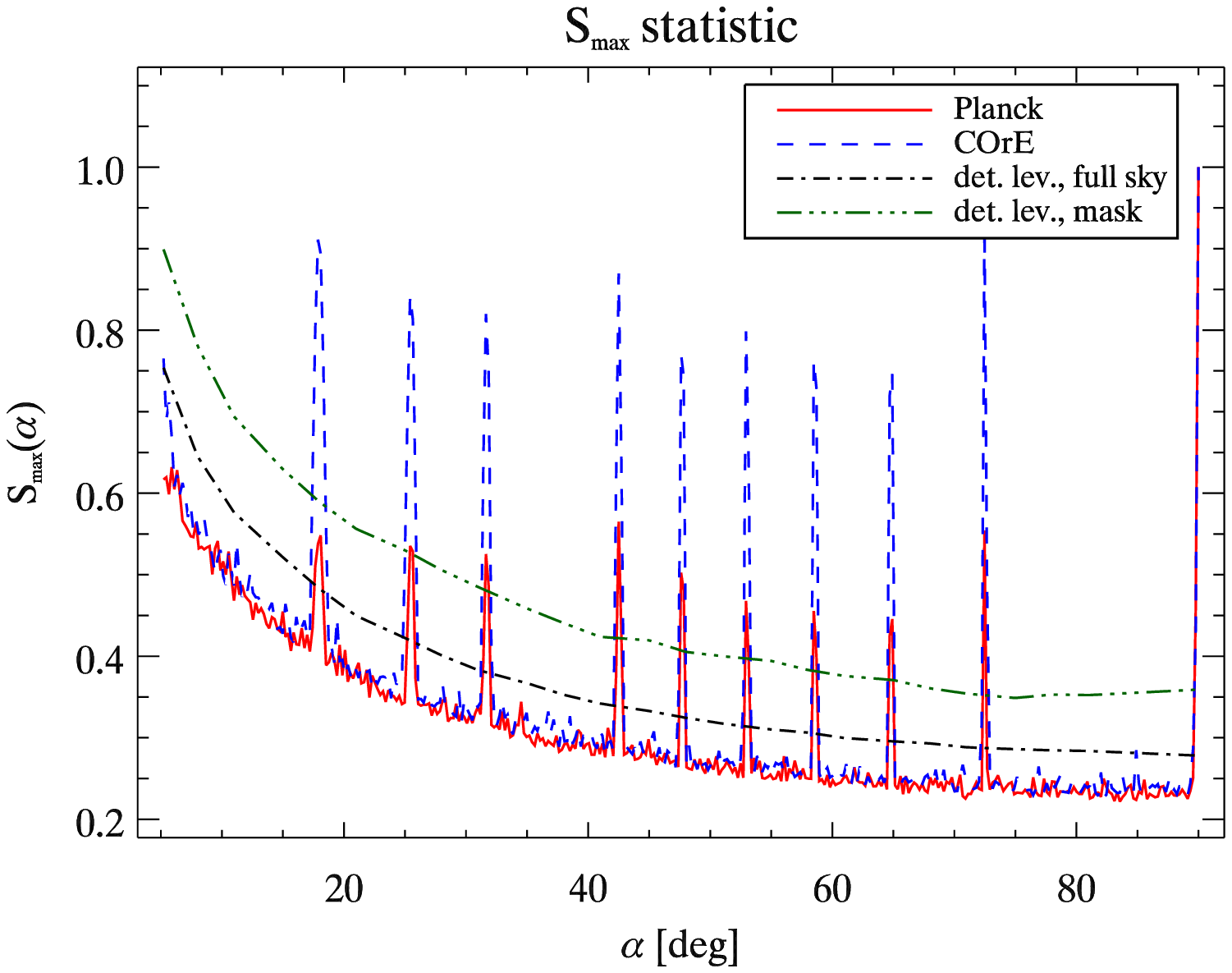,scale=0.35}
}
\caption{In the left figure, $S_{\rm max}^{\pm}$ statistic for the \emph{WMAP} 7-year W-band
  map. Solid and dotted lines show the statistics $S_{\rm max}^-$ and
  $S_{\rm max}^+$, respectively, for the W-band map masked with the
  KQ85y7 mask. The dashed line  
  is the false detection level estimated from 100 MC simulations.
  The peak at $90^\circ$ corresponds to a match
  between two copies of the same circle of radius $90^\circ$ centered
  around two antipodal points. In the middle and right figures,
  examples of the $S_{\rm max}^{-}$ statistics for simulated CMB
  temperature and polarisation anisotropy maps, respectively,  of
  universe with the topology of a cubic 3-torus with dimensions $L =
  2\ c/H_0$. In the middle figure, the dotted, solid, dashed and three dot-dashed lines show the statistic for
  CMB maps of the Sachs-Wolfe (SW) effect, the positive and negative correlations
  of the Doppler effect and total anisotropy, respectively. In the
  right figure, the solid and dashed lines show the statistic for simulated polarisation maps with angular resolution 
and noise level corresponding to the Planck and COrE data, respectively. The dot-dashed and three 
dot-dashed lines show the false detection levels for the statistic
estimated from 100 Monte Carlo simulations 
of the Planck coadded 100, 143 and 217~GHz frequency polarisation maps for the full sky and cut sky analysis, 
respectively.
\label{fig:smax}}
\end{figure}

In order to decrease the false detection level and be able to detect
matched circles with smaller radius, we analyzed the \emph{WMAP} data with the highest angular resolution i.e.\ the W-band
map, corrected for Galactic foregrounds and smoothed with a Gaussian
beam profile of the Full Width at Half Maximum (FWHM) $20'$ to decrease the noise level. 
The statistic for this map analysed with the KQ85y7 mask is shown in
\fig\ref{fig:smax}.

We did not find any statistically significant correlation of circle
pairs in the map. As shown in Bielewicz and Banday~\cite{bielewicz:2011}, the minimum radius at which the peaks expected for
the matching statistic are larger than the false detection level is about $\alpha_{\rm min} \approx
10^\circ$ for the W-band map. Thus, we can exclude any topology that predicts matching 
pairs of back-to-back circles larger than this radius. This implies
that in a flat universe described otherwise by the best-fit 7-year \emph{WMAP}
cosmological parameters, a lower bound on the size of the fundamental
domain is $d = 2\, R_{\rm LSS} \cos(\alpha_{\rm min}) \simeq 27.9\
\rm{Gpc}$, where $R_{\rm LSS}$ is the distance to the last scattering
surface. However, one has to keep in mind that this constraint concerns only those
universes with such dimensions and orientation of the fundamental domain
with respect to the mask that allow the detection of pairs of matched circles.

\section{Prospects of constraining the topology using CMB polarisation
maps} \label{sec:polarisation}
\subsection{Discussion of degrading effects for temperature maps} \label{sec:temp_degrade}
Since the signatures of topology are imprinted on the surface of last
scattering, any effects that dilute this image will also degrade
the ability to detect such signatures by means of the matched circles
statistic. In the case of the temperature fluctuations, there are two
sources of anisotropy generated at the
last scattering surface: the combination of the internal photon density
fluctuations and the Sachs-Wolfe (SW) effect and the Doppler effect.
In the latter case, the correlations for pairs of matched 
circles can be negative in universes with multi-connected topology. For pairs of
back-to-back circles with 
a radius smaller than 45$^\circ$ the Doppler term becomes increasingly
  anticorrelated. The consequences of these degrading effects are weaker constraints on the
topology of the Universe obtained from the matched circle
statistic. As we can see in \fig\ref{fig:smax}, use of the CMB map without both of the
degrading effects would allow us to impose lower bounds on the minimum radius of
the correlated circles which can be detected much below the present
constraints $\alpha_{\rm min} \approx 10^\circ$ (Bielewicz and Banday~\cite{bielewicz:2011}).

\subsection{Search of matched circles in polarisation maps} \label{sec:pol_search}
A polarisation map at small angular scales can be considered as a snapshot of
the last scattering surface. Theoretically, then, the polarisation provides a better
opportunity for the detection of multi-connected topology signatures
than a temperature anisotropy map. The only serious issues preventing
its use in studies of topology are instrumental noise and the
correction of the polarised data for the Galactic foreground.  

The $S^{-}_{\rm max}$ statistic for the simulated maps is
shown in \fig\ref{fig:smax}. As expected the amplitudes of the
peaks do not decrease with the radius of the circles as in the case of the
temperature anisotropy maps. We see that pairs of matched circles
can be detected for the Planck coadded 100, 143 and 217 GHz frequency
polarisation maps. However, comparing with figure for the temperature anisotropy
one should notice that the relative amplitude of the peaks with 
respect to the average correlation level for the circles with small radius is not
bigger than for the temperature map. Thus, the constraints on topology
will not be much tighter than those derived from an analysis of
temperature maps. A much better prospect arises for the COrE
maps. The signal of the multi-connected topologies is very pronounced
in this case enabling the detection of matched circles with very small
radius thus providing tighter constraints on the topology of the Universe.

\section*{Acknowledgments}
We acknowledge use of \textsc{camb}\footnote{http://camb.info/} \cite{camb}
and the \textsc{healpix}\footnote{http://healpix.jpl.nasa.gov}
software G\'orski {\it et al}.~\cite{gorski:2005}
and analysis package for deriving the results in this paper. We
acknowledge the use of the Legacy Archive for Microwave Background Data Analysis
(LAMBDA). Support for LAMBDA is provided by the NASA Office of Space
Science. The authors acknowledge the use of version 1.6.6 of the Planck
Sky Model, developed by the Component Separation Working Group (WG2)
of the \emph{Planck} Collaboration (Leach {\it et al}.~\cite{leach:2008},
Betoule {\it et al}.~\cite{betoule:2009}). Part of the computations were
performed at Interdisciplinary Center for Mathematical and
Computational Modeling at Warsaw University within a grant of
computing time no.\ G27-13.

\end{document}